\documentclass[sigplan,10pt]{acmart}

\pdfoutput=1  

\setcopyright{acmlicensed}
\acmPrice{15.00}
\acmDOI{10.1145/3497775.3503672}
\acmYear{2022}
\copyrightyear{2022}
\acmSubmissionID{poplws22cppmain-p3-p}
\acmISBN{978-1-4503-9182-5/22/01}
\acmConference[CPP '22]{Proceedings of the 11th ACM SIGPLAN International Conference on Certified Programs and Proofs}{January 17--18, 2022}{Philadelphia, PA, USA}
\acmBooktitle{Proceedings of the 11th ACM SIGPLAN International Conference on Certified Programs and Proofs (CPP '22), January 17--18, 2022, Philadelphia, PA, USA}



\usepackage{amsmath, amsthm, amsfonts, listings, MnSymbol, tikz}
\usepackage{hyperref}
\usepackage[utf8x]{inputenc}
\usepackage{upgreek}
\usepackage{xcolor}
\usepackage{colortbl}
\usepackage[inline]{enumitem}
\usepackage[capitalise]{cleveref}
\usepackage{subcaption}
\usepackage{tikz-cd}
\usepackage{tabularx}
\usepackage{listings}
\usepackage{flushend}
\usepackage{xspace}
\usepackage{underscore}
\usepackage{microtype}
\usepackage{scalefnt}

\makeatletter
\global\let\tikz@ensure@dollar@catcode=\relax
\makeatother

\definecolor{keywordcolor}{rgb}{0.7, 0.1, 0.1}   
\definecolor{tacticcolor}{rgb}{0.0, 0.1, 0.3}    
\definecolor{commentcolor}{rgb}{0.4, 0.4, 0.4}   
\definecolor{stringcolor}{rgb}{0.5, 0.3, 0.2}    
\definecolor{symbolcolor}{rgb}{0.1, 0.2, 0.7}    
\definecolor{sortcolor}{rgb}{0.1, 0.5, 0.1}      
\definecolor{attributecolor}{rgb}{0.7, 0.1, 0.1} 
\definecolor{errorcolor}{rgb}{1, 0, 0}           

\newcommand{\RR}{\mathbb{R}}

\numberwithin{equation}{section}

\DeclareMathOperator {\End}{End}

\DeclareMathOperator {\tr}{tr}
\DeclareMathOperator {\ad}{ad}

\newcommand{\extlink}[1]{
\href{https://github.com/leanprover-community/mathlib/blob/ba1cbfac3e0f2123bfa7fc13c7abcf0ff8002e4d/src/#1}
     {\def\svgwidth{0.8em}
\begingroup%
  \makeatletter%
  \providecommand\color[2][]{%
    \errmessage{(Inkscape) Color is used for the text in Inkscape, but the package 'color.sty' is not loaded}%
    \renewcommand\color[2][]{}%
  }%
  \providecommand\transparent[1]{%
    \errmessage{(Inkscape) Transparency is used (non-zero) for the text in Inkscape, but the package 'transparent.sty' is not loaded}%
    \renewcommand\transparent[1]{}%
  }%
  \providecommand\rotatebox[2]{#2}%
  \newcommand*\fsize{\dimexpr\f@size pt\relax}%
  \newcommand*\lineheight[1]{\fontsize{\fsize}{#1\fsize}\selectfont}%
  \ifx\svgwidth\undefined%
    \setlength{\unitlength}{12bp}%
    \ifx\svgscale\undefined%
      \relax%
    \else%
      \setlength{\unitlength}{\unitlength * \real{\svgscale}}%
    \fi%
  \else%
    \setlength{\unitlength}{\svgwidth}%
  \fi%
  \global\let\svgwidth\undefined%
  \global\let\svgscale\undefined%
  \makeatother%
  \begin{picture}(1,1)%
    \lineheight{1}%
    \setlength\tabcolsep{0pt}%
    \put(0,0){\includegraphics[width=\unitlength,page=1]{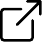}}%
  \end{picture}%
\endgroup%
}
}

\begin{document}

\title{Formalising Lie algebras}

\author{Oliver Nash}
\email{acmsigplan@olivernash.org}
\orcid{}

\affiliation{
  \institution{Imperial College}
  \streetaddress{South Kensington}
  \postcode{SW7 2AZ}
  \city{London}
  \country{United Kingdom}
}

\begin{abstract}
  Lie algebras are an important class of algebras which arise throughout mathematics and physics.
  We report on the formalisation of Lie algebras in Lean's Mathlib library.
  Although basic knowledge of Lie theory will benefit the reader, none is assumed; the
  intention is that the overall themes will be accessible even to readers unfamiliar with
  Lie theory.

  Particular attention is paid to the construction of the classical and exceptional Lie algebras.
  Thanks to these constructions, it is possible to state the classification theorem for
  finite-dimensional semisimple Lie algebras over an algebraically closed field of characteristic
  zero.

  In addition to the focus on Lie theory, we also aim to highlight the unity of Mathlib.
  To this end, we include examples of achievements made possible only by leaning
  on several branches of the library simultaneously.
\end{abstract}

\maketitle

\section{Introduction}
\subsection{Skew-symmetric matrices}\label{skewSymmSubsect}
Recall that a rotation in Euclidean space can be represented by an invertible matrix $X$
whose inverse is its transpose:
\begin{align*}
  X^{-1} = X^T.
\end{align*}

At least as far back at the 19$^{\rm th}$ Century, it was observed that if $A$ is a skew-symmetric
matrix and $\epsilon \in \RR$ is small then $I + \epsilon A$ is almost a rotation. Indeed
since $A^T = -A$, we have:
\begin{align*}
  (I + \epsilon A)^T (I + \epsilon A) = I - \epsilon^2 A^2.
\end{align*}
That is, the inverse of $I + \epsilon A$ is its transpose, if we neglect terms order $\epsilon^2$.

Better yet, the exponential $e^{\epsilon A}$ is truly a rotation (no terms neglected) and
for another such matrix $B$, the Baker-Campbell-Hausdorff formula quantifies how the composition
of the rotations $e^{\epsilon A}$, $e^{\epsilon B}$ behaves in terms of skew-symmetric matrices:
\begin{align}\label{CBH}
  e^{\epsilon A}e^{\epsilon B} = e^{\epsilon (A + B) + \frac{\epsilon^2}{2}[A, B] + O(\epsilon^3)}.
\end{align}
The term $[A, B]$ appearing in \eqref{CBH} is defined as:
\begin{align}\label{LBdef}
  [A, B] = AB - BA
\end{align}
and is an instance of a Lie bracket. It defines a natural product:
if $A$ and $B$ are skew-symmetric then\footnote{If this is the first time you have seen
this, then check: it's a fun calculation.} so is $[A, B]$.

The Lie bracket is skew-commutative:
\begin{align*}
  [A, B] = -[B, A],
\end{align*}
and in general non-associative,
but by way of compensation it satisfies the Jacobi identity:
\begin{align}\label{Jacobi}
  [A, [B, C]] + [B, [C, A]] + [C, [A, B]] = 0.
\end{align}

\subsection{Abstract Lie algebras and their ubiquity}\label{ubiquitySubsect}
Recognising skew-symmetric matrices merely as an example, one can consider the study of
abstract Lie algebras. These are modules carrying a bilinear, skew-commutative product which
satisfies the Jacobi identity \eqref{Jacobi}.

The study of abstract Lie algebras was initiated by Lie and independently by Killing
more than 140 years ago \cite{MR1510035}, \cite{MR1510482,MR1510529,MR1510568}, \cite{MR1007036}
and the subject now pervades much of modern mathematics and physics.

In classical physics, both linear and angular momentum are best understood as taking values
in the dual of a Lie algebra. More importantly, recognising that the space of classical observables
forms a Lie algebra\footnote{In fact, a Poisson algebra, but it is the bracket structure that
is more subtle.} under the Poisson bracket is an important
step in quantisation. Furthermore, in particle physics, elementary particles such as quarks
are essentially basis vectors of irreducible representations of Lie groups \cite{MR2651086} and
thus of Lie algebras.

In number theory, automorphic forms, central objects of study in the Langlands programme,
satisfy a differential equation defined in terms of a reductive Lie algebra. In
differential geometry, the tangent bundle is special amongst
vector bundles because its sections carry a natural Lie algebra structure; moreover
with just this structure one can define the de Rham cohomology, thus connecting with algebraic
topology. In Riemannian geometry
and gauge theory, the curvature 2-form takes values in a Lie algebra. In symplectic geometry, the
moment map takes values in the dual of a Lie algebra.

Of course the lists above hardly scratch the surface. The unifying theme is that a great many types
of symmetry are naturally Lie groups or algebraic groups,
and thus have associated Lie algebras which are essential for their study. Understanding symmetry
thus requires understanding Lie algebras and for this reason
the classification of semisimple Lie algebras is rightly regarded as a landmark result, obtained
just in time for the 20$^{\rm th}$ Century\footnote{Cartan submitted his thesis
\cite{Cartan1894} in March 1894.}.

\subsection{A roadmap for this article}
Much of the work discussed here was motivated by the desire to formalise the statement
of the classification of semisimple Lie algebras in a proof assistant. This statement
appears in section \ref{classificationSect} of this article and the intervening sections
\ref{designChoiceSect} -- \ref{exceptionalSect} essentially correspond to the various
waypoints which were necessarily passed on the way to this milestone.

Aside from section \ref{designChoiceSect}, which is included to give a taste for
foundational design decisions, these intervening sections fall neatly into two groups.

The first group, which consists of sections \ref{idealOperSect}, \ref{radicalSect},
enables us to define the class of Lie algebras we are classifying.

The second group, which
consists of sections \ref{ofAssocSect} -- \ref{exceptionalSect}, enables us to
define the concrete Lie algebras which, according to the classification theorem, exhaust the
class of semisimple Lie algebras (up to equivalence).

Section \ref{weightRootSect} sketches the formalisation of weights and roots in Lie
theory, and section \ref{finalWordsSect} concludes with some general remarks.

\subsection{A primer on Lean and Mathlib}
The work dicussed here was implemented using Lean\footnote{More precisely the
\href{https://github.com/leanprover-community/lean}{community fork} of Lean 3.}.
Lean is a dependently-typed programming language together with a proof assistant.

Like Coq, Lean is based on the Calculus of Inductive Constructions; see \cite{CarneiroMaster}
for a detailed discussion. We do not expect the
reader to be an expert in Lean. For the purposes of this article, an intuitive
understanding of the following Lean keywords should suffice:
\begin{itemize}
  \item \texttt{variables}
  \item \texttt{def}
  \item \texttt{abbreviation}
  \item \texttt{lemma}
  \item \texttt{class}
  \item \texttt{instance}
\end{itemize}
The keyword \texttt{variables} adds variables to the local context; \texttt{def} and
its variant \texttt{abbreviation} make definitions. The keyword \texttt{lemma} is
self-explanatory but \texttt{class} and its partner \texttt{instance} deserve
further comment.

The keyword \texttt{class} defines typeclasses.
Lean includes a powerful typeclass system which is heavily used in
its Mathlib \cite{Mathlib} library. For
example Mathlib contains the typeclass \texttt{comm_ring} which defines what it means
for a type to carry the structure of a commutative ring (with unit). Using this, we can
say that a type \texttt{R} is a commutative ring by supplying a typeclass argument
\texttt{[comm_ring R]} in the statement of definition or lemma.

Typeclasses have
one constructor which may take several arguments. In the case of a commutative ring,
the arguments correspond to an addition function, a multiplication function,
associativity of addition, associativity of multiplication, commutativity of
multiplication etc.

Lastly, the keyword \texttt{instance} is what makes typeclasses so useful: it allows
us to register the fact that some type carries a typeclass. These instance statements
can even contain mathematically non-trivial facts. For example here is Mathlib's
statement that the real numbers form a commutative ring\footnote{Notice the
\extlink{data/real/basic.lean\#L65} icon;
it is a permalink to the corresponding code in Mathlib. We provide such permalinks
throughout the text so that readers using an internet-connected device may easily
navigate to appropriate locations in Mathlib.} \extlink{data/real/basic.lean\#L65}:
\begin{lstlisting}
instance : comm_ring ℝ :=
begin
  -- proof using Cauchy sequences (omitted here)
end
\end{lstlisting}
This instance means Lean knows that any lemmas about commutative rings automatically
hold for the real numbers\footnote{Of course more is true, e.g., the real numbers
are a linearly-ordered field and Mathlib knows this too.}.

Aside from \texttt{comm_ring}, the most important typeclasses expliclty used in the work discussed
here are \texttt{add_comm_group}, which states that a type carries the structure of a
commutative group (with group operation denoted \texttt{+}) and \texttt{module}, which
states that a type carries the structure of a module over a set of scalars. For readers
not familiar with modules, simply read {\lq}vector space{\rq} instead.

Finally, since of all the work discussed here has been merged to the Mathlib master branch,
anyone wishing to
run, compile, interact with, or build upon Mathlib's Lie algebras can do so by following instructions
available at the Lean community website, especially:
\begin{center}
\href{https://leanprover-community.github.io/get_started.html}
     {https://leanprover-community.github.io/get_started.html}
\end{center}
and:
\begin{center}
\href{https://leanprover-community.github.io/leanproject.html}
     {https://leanprover-community.github.io/leanproject.html}.
\end{center}
Online documentation is automatically generated for all Mathlib code. For
Lie algebras, a good entry point is:
\begin{center}
\href{https://leanprover-community.github.io/mathlib_docs/algebra/lie/basic.html}
     {https://leanprover-community.github.io/mathlib_docs/algebra/lie/basic.html}.
\end{center}

\subsection{Lie algebras in Lean}\label{definitionSubSect}
Here is the definition of a Lie algebra in Mathlib\extlink{algebra/lie/basic.lean\#L61}:
\begin{lstlisting}
class lie_ring (L : Type v)
  extends add_comm_group L, has_bracket L L :=
(add_lie : ∀ (x y z : L),
  ⁅x + y,z⁆ = ⁅x,z⁆ + ⁅y,z⁆)
(lie_add : ∀ (x y z : L),
  ⁅x,y + z⁆ = ⁅x,y⁆ + ⁅x,z⁆)
(lie_self : ∀ (x : L), ⁅x,x⁆ = 0)
(leibniz_lie : ∀ (x y z : L),
  ⁅x,⁅y,z⁆⁆ = ⁅⁅x,y⁆,z⁆ + ⁅y,⁅x,z⁆⁆)

class lie_algebra (R : Type u) (L : Type v)
  [comm_ring R] [lie_ring L] extends module R L :=
(lie_smul : ∀ (t : R) (x y : L),
  ⁅x,t • y⁆ = t • ⁅x,y⁆)
\end{lstlisting}
The skew-commutative property follows from the \texttt{lie_self} axiom and the Jacobi identity
is equivalent to the \texttt{leibniz_lie} axiom\footnote{We comment on the choice of axioms in
section \ref{designChoiceSect}.}.

There exist computer algebra systems such as SageMath, GAP, MAGMA as well as a Mathematica
package available at \href{http://katlas.org}{http://katlas.org}, and the Lie-specific package
LiE \cite{lieSoftware2000}, that are capable of performing calculations involving Lie algebras.
However to the best of our knowledge, there is no previous work formalising the theory of Lie
algebras.

As of December 2021, Mathlib contains over 6,000 lines of code
about Lie algebras and their representations, broadly following Bourbaki
\cite{bourbaki1975, bourbaki1968, bourbaki1975b}. Material covered includes \extlink{algebra/lie}:
\begin{itemize}
  \item Lie algebras and Lie modules
  \item Morphisms and equivalences of Lie algebras and Lie modules
  \item Lie subalgebras, Lie submodules, Lie ideals, and quotients
  \item Extension and restriction of scalars
  \item Direct sums of Lie modules and Lie algebras
  \item Tensor product of Lie modules
  \item Lie ideal operations, the lower central series, the derived series, and derived length
  \item Nilpotent, solvable, simple, semisimple Lie algebras, the radical, and the
  centre of a Lie algebra
  \item Cartan subalgebras
  \item Weight spaces of a Lie module, and thus root spaces of a Lie algebra
  \item The universal enveloping algebra (and its universal property)
  \item The free Lie algebra (and its universal property)
  \item Definition of the classical Lie algebras
  \item Definition of the exceptional Lie algebras
\end{itemize}
The final item is worth highlighting. There is no easy route to the definition of the exceptional
Lie algebras (section \ref{exceptionalSect}) and it is an important milestone since it allows us
to state the classification of semisimple Lie algebras (section \ref{classificationSect}).
A \emph{proof} of this classification within Mathlib would be a significant undertaking but
now looks achievable.

\subsection{A note about notation}
We draw the reader's attention to the brackets appearing in Lean code such as the equation:
\begin{lstlisting}
  ⁅x,⁅y,z⁆⁆ = ⁅⁅x,y⁆,z⁆ + ⁅y,⁅x,z⁆⁆
\end{lstlisting}
appearing above. These brackets, associated with the \texttt{has_bracket}
typeclass\extlink{data/bracket.lean\#L37}, were
introduced to Mathlib to provide a convenient notation for the Lie bracket\footnote{Actually
they have since been used elsewhere, e.g., for the commutator of two subgroups of a group.}.

They are also used in the notation for morphisms in Lie theory. For example, the
following is Mathlib's notation for an $R$-linear map of modules:
\begin{lstlisting}
  M₁ →ₗ[R] M₂
\end{lstlisting}
whereas the following is the notation for a morphism of Lie algebras:
\begin{lstlisting}
  L₁ →ₗ⁅R⁆ L₂
\end{lstlisting}
and the following is the notation for a morphism of Lie modules over a Lie algebra $L$
with coefficients in $R$:
\begin{lstlisting}
  M₁ →ₗ⁅R,L⁆ M₂
\end{lstlisting}

Finally, similar remarks apply to equivalences, i.e., we use the notations:
\begin{lstlisting}
  L₁ ≃ₗ⁅R⁆ L₂
\end{lstlisting}
and:
\begin{lstlisting}
  M₁ ≃ₗ⁅R,L⁆ M₂
\end{lstlisting}

\section{Design choices: Leibniz vs. Jacobi}\label{designChoiceSect}
The choice of the axiom \texttt{leibniz_lie} in the definition exhibited in
section \ref{definitionSubSect} deserves explanation, if only because it serves as a
simple example of the sorts of choices that repeatedly came up
in the course of formalisation.

In the presence of the other Lie algebra axioms, each of the following are equivalent:
\begin{align}
  [x, [y, z]] + [y, [z, x]] + [z, [x, y]] = 0\label{jacobiAxiom},\\
  [[x, y], z] = [x, [y, z]] - [y, [x, z]]\label{lieLieAxiom},\\
  [x, [y, z]] = [[x, y], z] + [y, [x, z]]\label{leibnizLieAxiom}.
\end{align}
Note that \eqref{leibnizLieAxiom} is the axiom \texttt{leibniz_lie} and that
given $x : L$, if we define $D_x : L \to L$ by:
\begin{align*}
  D_x y = [x, y]
\end{align*}
then \eqref{leibnizLieAxiom} says that $D_x$ satisfies the Leibniz product rule:
\begin{align*}
  D_x [y, z] = [D_x y, z] + [y, D_x z],
\end{align*}
i.e., it says that $D_x$ is a derivation.

One might think that the Jacobi identity \eqref{jacobiAxiom} is the best choice since it looks
the most symmetric; in fact it is the worst choice. This becomes clear when one considers that
we also need a theory of Lie modules.

Informally, if a Lie algebra $L$ acts linearly on a module
$M$, and if we denote the action of $x : L$ on $m : M$ by $[x, m] : M$ then this action turns $M$
into a Lie module for $L$ iff:
\begin{align*}
  [x, [y, m]] = [[x, y], m] + [y, [x, m]],
\end{align*}
for all $x, y : L$ and $m : M$.

Now consider the case $L = M$ and observe that any Lie algebra is thus a module over itself. This
so-called adjoint action is extremely important in Lie theory. Observe also that if we replace
$z : L$ in equations \eqref{jacobiAxiom} -- \eqref{leibnizLieAxiom} by $m : M$ then the terms
$[y, [z, x]]$, $[z, [x, y]]$ do not make sense since there is no action of $M$ on $L$. Thus
\eqref{jacobiAxiom} cannot be used as an axiom for Lie modules.

By choosing to define Lie algebras
using the \texttt{leibniz_lie} axiom \eqref{leibnizLieAxiom} we thus obtain a theory where Lie
algebras and Lie modules \emph{definitionally} satisfy the same axiom. This is a desirable
convenience that we exploit. For example, here is the code that defines the adjoint action
\extlink{algebra/lie/basic.lean\#L107}:
\begin{lstlisting}
instance lie_self_module : lie_ring_module L L :=
{ .. (infer_instance : lie_ring L) }
\end{lstlisting}

One might ask why to choose \eqref{leibnizLieAxiom} over \eqref{lieLieAxiom}. This is of lesser
importance but \eqref{leibnizLieAxiom} is still the better choice. This is because
\eqref{lieLieAxiom} requires using subtraction which is often a secondary
operation, defined via addition and inverses. This means that when constructing Lie algebras
downstream, it is likely there will be more direct proof of \eqref{leibnizLieAxiom}.

On the other hand, as a simplification lemma (rather than a definition) \eqref{lieLieAxiom} is
excellent since it can be used to push Lie brackets right-most in nested expressions.
Indeed the following \texttt{simp} lemma establishes this as the normal form
in Mathlib\footnote{Here and elsewhere we have omitted the proof. In the
actual code (available via the permalinks) the proof follows immediately after the
\texttt{:=} syntax.}
\extlink{algebra/lie/basic.lean\#L142}:
\begin{lstlisting}
@[simp] lemma lie_lie :
  ⁅⁅x,y⁆,m⁆ = ⁅x,⁅y,m⁆⁆ - ⁅y,⁅x,m⁆⁆ :=
\end{lstlisting}
And of course we could never register \eqref{leibnizLieAxiom} as \texttt{simp} lemma since we
would get a \texttt{simp} loop: the term $[y, [x, m]]$ on the right hand side of
\eqref{leibnizLieAxiom} is of the same form as the term $[x, [y, m]]$ on the left.

Notwithstanding the words above, the choice of axiom here is of minor importance. However
there are many such choices and en mass they accumulate to have non-trivial impact.

\section{Ideal operations, solvable Lie algebras, nilpotent Lie modules}\label{idealOperSect}
Several important constructions in Lie theory are conveniently stated in the language of ideal
operations.

Informally, if $L$ is a Lie algebra, $M$ is a Lie module of $L$, $I$ is an ideal of $L$, and $N$
is a Lie submodule of $M$ we define:
\begin{align*}
  [I, N] = ~&\mbox{smallest Lie submodule of $M$ containing}\\
           ~&\mbox{$[x, n]$ for all $x : I$, $n : N$.}
\end{align*}
Note that since any Lie algebra can be regarded as a Lie module over itself, and since ideals
are just Lie submodules, we can thus combine two ideals $I$, $J$ to make another: $[I, J]$.

Taking advantage of the complete lattice structure on Lie submodules, we formalised the above as
\extlink{algebra/lie/submodule.lean\#L404}:
\begin{lstlisting}
def lie_span (s : set M) : lie_submodule R L M :=
Inf {N | s ⊆ N}
\end{lstlisting}
and \extlink{algebra/lie/ideal_operations.lean\#L47}:
\begin{lstlisting}
instance : has_bracket
  (lie_ideal R L) (lie_submodule R L M) :=
⟨λ I N, lie_span R L
  {m | ∃(x : I) (n : N), ⁅(x : L),(n : M)⁆ = m}⟩
\end{lstlisting}
This definition is compatible with the lattice structure on Lie submodules in numerous ways.
For example \extlink{algebra/lie/ideal_operations.lean\#L89}:
\begin{lstlisting}
lemma lie_le_right : ⁅I, N⁆ ≤ N :=

lemma lie_comm : ⁅I, J⁆ = ⁅J, I⁆ :=

lemma lie_le_inf : ⁅I, J⁆ ≤ I ⊓ J :=

@[simp] lemma lie_sup :
  ⁅I, N ⊔ N'⁆ = ⁅I, N⁆ ⊔ ⁅I, N'⁆ :=

lemma mono_lie (h₁ : I ≤ J) (h₂ : N ≤ N') :
  ⁅I, N⁆ ≤ ⁅J, N'⁆ :=
\end{lstlisting}

We also established alternate characterisations of the definition $[I, N]$.
Firstly \extlink{algebra/lie/ideal_operations.lean\#L54}:
\begin{lstlisting}
lemma lie_ideal_oper_eq_linear_span :
  (↑⁅I, N⁆ : submodule R M) = submodule.span R
  {m | ∃(x : I) (n : N), ⁅(x : L),(n : M)⁆ = m} :=
\end{lstlisting}
This says that if we forget the action of $L$ and regard $[I, N]$ merely as a submodule of $M$
then it is just the \emph{linear} span of the generating elements (rather than the Lie span).
This result was very useful.

Secondly, we established a characterisation that does not use spans at all
\extlink{algebra/lie/tensor_product.lean\#L191}:
\begin{lstlisting}
lemma lie_ideal_oper_eq_tensor_map_range :
  ⁅I, N⁆ = ((to_module_hom R L M).comp
  (map_incl I N : ↥I ⊗ ↥N→ₗ⁅R,L⁆ L ⊗ M)).range :=
\end{lstlisting}
Informally, from the data $I$, $N$ we can build a composition of morphisms of Lie modules:
\begin{align*}
  I ⊗ N → L ⊗ M → M,
\end{align*}
where the first arrow is the tensor product of inclusion maps and the second arrow is the action
of $L$. The lemma states that the range of this composite map is $[I, N]$.

Probably the most important application of these ideal operations are the definitions of
the derived series and lower central series.

Informally the derived series, $D^0 L, D^1 L, D^2 L, \ldots $ of a Lie algebra is a sequence of
ideals of $L$, defined inductively using the ideal operation discussed above:
\begin{align*}
  D^0 L     &= L\\
  D^{k+1} L &= [D^k L, D^k L].
\end{align*}
We formalised the derived series as \extlink{algebra/lie/solvable.lean\#L50}:
\begin{lstlisting}
def derived_series_of_ideal (k : ℕ) :
  lie_ideal R L → lie_ideal R L :=
(λ I, ⁅I, I⁆)^[k]

abbreviation derived_series (k : ℕ) :
  lie_ideal R L :=
derived_series_of_ideal R L k ⊤
\end{lstlisting}
Note that we defined \texttt{derived_series} as a special case of a more general definition
\texttt{derived_series_of_ideal}. This was very useful since it provides a type-theoretic
expression of the fact that if we regard a Lie ideal as a Lie algebra in its own right, then the
terms of its derived series are also ideals of the enclosing algebra\footnote{Just like with normal
subgroups of a group, if $I$ is an ideal of a Lie algebra $L$ and $J$ is an ideal of $I$ it is not
necessarily true that $J$ is an ideal of $L$.}. Here is the statement that the two concepts
really do agree when we regard an ideal as a Lie algebra in its own right \extlink{algebra/lie/solvable.lean\#L133}:
\begin{lstlisting}
lemma
  derived_series_eq_derived_series_of_ideal_comap
  (k : ℕ) :
  derived_series R I k =
  (derived_series_of_ideal R L k I).comap I.incl:=
\end{lstlisting}

We then used the derived series to define what it means for a Lie algebra to be
solvable \extlink{algebra/lie/solvable.lean\#L186}:
\begin{lstlisting}
class is_solvable : Prop :=
(solvable : ∃ k, derived_series R L k = ⊥)
\end{lstlisting}
and built out the standard theory.

By way of example, we recall that the standard example of a solvable Lie
algebra is the set of upper-triangular square matrices\footnote{Again, if this is the first
time you've seen this, it is a fun calculation to verify that if $A$, $B$ are upper-triangular
matrices then so are $AB$ and $BA$, and thus also $[A, B]$.}.

Similarly, we defined the lower central series \extlink{algebra/lie/nilpotent.lean\#L35},
and used it to define the concept of nilpotency \extlink{algebra/lie/nilpotent.lean\#L76}.
In this case we generalised slightly from the standard references since the concept of
nilpotency makes sense not just for Lie algebras but for Lie modules and so we made this
more general definition. This turned out to be useful when formalising Engel's theorem (to appear).

\section{Case study: the radical is solvable}\label{radicalSect}
Whenever possible, we strove to work at the greatest reasonable level of generality. At times
the unified nature of Mathlib made it possible to establish results at a level of generality beyond
that of the standard references, including Bourbaki. A good example is the basic result that
finite-dimensional Lie algebras possess a maximal solvable ideal.

As we have seen, Lie algebras admit a notion of being solvable. For the purposes of this
discussion, the precise meaning is unimportant. What is important is that if $I$, $J$ are ideals
of a Lie algebra and if, regarding them as Lie algebras in their own right, they are
both solvable, then their sum $I + J$ is solvable. Mathlib knows this fact. Indeed here is the
statement and proof for a Lie algebra $L$ over a commutative ring $R$ \extlink{algebra/lie/solvable.lean\#L192}:
\begin{lstlisting}
instance is_solvable_add {I J : lie_ideal R L}
  [hI : is_solvable R I] [hJ : is_solvable R J] :
  is_solvable R ↥(I + J) :=
begin
  tactic.unfreeze_local_instances,
  obtain ⟨k, hk⟩ := hI,
  obtain ⟨l, hl⟩ := hJ,
  exact ⟨⟨k+l,
    lie_ideal.derived_series_add_eq_bot hk hl⟩⟩,
end
\end{lstlisting}

The (solvable) radical of a Lie algebra is the sum of all solvable ideals, or more precisely, the
supremum of the subset of solvable ideals in the complete lattice of ideals of a Lie algebra. Here is
the definition in Mathlib \extlink{algebra/lie/solvable.lean\#L257}:
\begin{lstlisting}
def radical :=
Sup { I : lie_ideal R L | is_solvable R I }
\end{lstlisting}

It is clear that if $R$ is a field and $L$ is finite-dimensional then the radical itself is
finite-dimensional and can thus be represented as a sum of finitely-many solvable ideals. By
iterating \texttt{is_solvable_add} we thus see the radical is solvable. This was the greatest
level of generality in which this fact was established in any reference the author could find.

However it is not necessary to make such strong assumptions. Indeed the result is true over
any commutative ring $R$ as long as $L$ is Noetherian, as can be seen from the following proof
in Mathlib \extlink{algebra/lie/solvable.lean\#L260}:
\begin{lstlisting}
instance radical_is_solvable [is_noetherian R L] :
  is_solvable R (radical R L) :=
begin
  have h := lie_submodule.
    well_founded_of_noetherian R L L,
  rw ← complete_lattice.
    is_sup_closed_compact_iff_well_founded at h,
  refine
    h { I : lie_ideal R L | is_solvable R I } _ _,
  { use ⊥,
    exact lie_algebra.is_solvable_bot R L, },
  { intros I J hI hJ,
    apply lie_algebra.is_solvable_add R L;
    [exact hI, exact hJ], },
end
\end{lstlisting}

The key lemma is \texttt{complete_lattice.is_sup_closed_compact_iff_well_founded}
\extlink{order/compactly_generated.lean\#L223} which
the author added to the lattice theory library for the purposes of proving
\texttt{radical_is_solvable}. This addition was only possible because Mathlib already contained
a comprehensive lattice theory library and numerous key results about well-founded relations.
Furthermore the lemma \texttt{lie_submodule.well_founded_of_noetherian}
\extlink{algebra/lie/submodule.lean\#L347} ultimately depends
upon results which were originally introduced to Mathlib for the purposes of formalising
results about Noetherian modules over commutative rings with a view toward algebraic geometry.

Different people with different aims in different corners of Mathlib are enabling each other to
push boundaries into new territory.

\section{Lie algebras from associative algebras}\label{ofAssocSect}
Any associative algebra $A$ carries a natural Lie algebra structure via the
definition:
\begin{align*}
  [x, y] = xy - yx.
\end{align*}
This is an extremely important\footnote{Indeed any Lie algebra that injects
into its universal enveloping algebra is a Lie subalgebra of such a Lie algebra, or better yet
(in finite dimensions) see Ado's theorem \cite{bourbaki1975} I §7.3.}
class of Lie algebras which we needed early on.

We thus registered the following data-bearing typeclass instance
\extlink{algebra/lie/of_associative.lean\#L45}:
\begin{lstlisting}
instance {A : Type*} [ring A] : has_bracket A A :=
⟨λ x y, x*y - y*x⟩
\end{lstlisting}
together with instances containing proofs that this definition satisfies the required
axioms \extlink{algebra/lie/of_associative.lean\#L55}:
\begin{lstlisting}
instance {A : Type*} [ring A] : lie_ring A :=
\end{lstlisting}
and \extlink{algebra/lie/of_associative.lean\#L76}:
\begin{lstlisting}
instance {R A : Type*} [comm_ring R] [ring A]
  [algebra R A] : lie_algebra R A :=
\end{lstlisting}
We also established basic properties about this correspondence. In particular we needed
to establish that a morphism of associative algebras can be regarded as a morphism of
Lie algebras \extlink{algebra/lie/of_associative.lean\#L88}:
\begin{lstlisting}
def alg_hom.to_lie_hom (f : A →ₐ[R] B) :
  A →ₗ⁅R⁆ B :=
\end{lstlisting}
All of the above followed easily using standard tactics.

\section{Skew adjoint endomorphisms}\label{skewAdjointSect}
If an associative algebra $A$ carries appropriate additional structure, it contains distinguished Lie
subalgebras when regarded as a Lie algebra in the sense of section \ref{ofAssocSect}.

The most
important examples of this phenomenon occur when $A$ is the endomorphisms of a module $M$, i.e.
$A = \End(M)$. These include:
\begin{enumerate}
  \item If $M$ is free with finite rank, $A$ contains the distinguished Lie subalgebra of trace zero
  elements.
  \item If $M$ carries a bilinear form, $A$ contains the distinguished Lie subalgebra of
  skew-adjoint elements.
  \item If $M$ carries a bilinear multiplication, $A$ contains the distinguished Lie subalgebra of
  derivations.
  \item If $M$ carries both a bilinear form and a compatible bilinear multiplication, $A$ contains
  the distinguished Lie subalgebra of skew-adjoint derivations.
\end{enumerate}

Our focus here is the second item above: the Lie subalgebra of skew-adjoint endomorphisms obtained
from a module carrying a bilinear form.

If the bilinear form is the dot product,
the skew-adjoint endomorphisms are just the skew-symmetric matrices of section \ref{skewSymmSubsect},
but as we shall see, by allowing more general bilinear forms, we obtain more general Lie algebras.

Informally, given $R$-modules $M$, $M'$ carrying bilinear forms:
\begin{align*}
  B : M &\times M \to R,\\
  B' : M' &\times M' \to R,
\end{align*}
we say that linear maps $f : M \to M'$ and $g : M' \to M$ are
adjoint\footnote{More precisely, $f$ is left adjoint to $g$.} iff:
\begin{align*}
  B'(f x, y) = B(x, g y),
\end{align*}
for all $x : M$, $y : M'$. Building on top of the existing theory of bilinear forms, we formalised
this as follows \extlink{linear_algebra/bilinear_form.lean\#L1040}:
\begin{lstlisting}
def bilin_form.is_adjoint_pair :=
∀ ⦃x y⦄, B' (f x) y = B x (g y)
\end{lstlisting}
In the special case $M = M'$ and $B = B'$ we say that $f$ is self-adjoint if:
\begin{align*}
  B(f x, y) = B(x, f y),
\end{align*}
for all $x, y : M$ and we say $f$ is skew-adjoint if:
\begin{align*}
  B(f x, y) = -B(x, f y),
\end{align*}
for all $x, y : M$.

We formalised these concepts in Mathlib as \extlink{linear_algebra/bilinear_form.lean\#L1129}:
\begin{lstlisting}
def is_self_adjoint := is_adjoint_pair B B f f
\end{lstlisting}
and \extlink{linear_algebra/bilinear_form.lean\#L1133}:
\begin{lstlisting}
def is_skew_adjoint := is_adjoint_pair B B f (-f)
\end{lstlisting}
and proved that the subsets of self-adjoint
and skew-adjoint endomorphisms are both submodules of $\End(M)$. To avoid code duplication
we introduced the concept of {\lq}pair-self-adjointness{\rq} \extlink{linear_algebra/bilinear_form.lean\#L1097}:
\begin{lstlisting}
def is_pair_self_adjoint :=
is_adjoint_pair B B' f f
\end{lstlisting}
where we remain specialised to a single module $M = M'$ but bring back the second
bilinear form $B'$.

When a module carries a single bilinear form $B$, the usual concept of
self-adjointness is pair-self-adjointness for the pair of bilinear forms $(B, B)$ and the usual
concept of skew-adjointness is pair-self-adjointness for the pair of bilinear forms $(-B, B)$.
The relevant formal statement is \extlink{linear_algebra/bilinear_form.lean\#L1136}:
\begin{lstlisting}
lemma is_skew_adjoint_iff_neg_self_adjoint :
  B.is_skew_adjoint f ↔
  is_adjoint_pair (-B) B f f :=
\end{lstlisting}
We then proved that for any pair of bilinear forms, the subset of pair-self-adjoint
endomorphisms forms a submodule of $\End(M)$.

Restricting our attention to just the skew-adjoint endomorphisms\footnote{In fact the subset of
self-adjoint endomorphisms also carries some extra structure: they form a Jordan subalgebra under
the product $x \circ y = xy + yx$.}, we then proved \extlink{algebra/lie/skew_adjoint.lean\#L41}:
\begin{lstlisting}
lemma bilin_form.is_skew_adjoint_bracket
  (f g : module.End R M)
  (hf : f ∈ B.skew_adjoint_submodule)
  (hg : g ∈ B.skew_adjoint_submodule) :
  ⁅f, g⁆ ∈ B.skew_adjoint_submodule :=
\end{lstlisting}
and so deduced that they form a Lie subalgebra, as required.

Finally, recalling that a square matrix $J$ defines a bilinear form on vectors:
\begin{align*}
  (v, w) \mapsto v^T J w,
\end{align*}
and that another square matrix $A$ defines an endomorphism of vectors:
\begin{align*}
  v \mapsto A v,
\end{align*}
we introduced the concept of adjointness for matrices. This turns out to
be \extlink{linear_algebra/bilinear_form.lean\#L1168}:
\begin{lstlisting}
def matrix.is_adjoint_pair := Aᵀ ⬝ J' = J ⬝ B
\end{lstlisting}
We constructed an API for matrices similar to the one for bilinear forms and proved that the
notions of adjointness correspond:
\begin{lstlisting}
lemma matrix_is_adjoint_pair_bilin_form :
  matrix.is_adjoint_pair J J' A B ↔
  bilin_form.is_adjoint_pair
    J.to_bilin_form J'.to_bilin_form
    A.to_lin B.to_lin :=
\end{lstlisting}

\section{The classical Lie algebras}
We made an early effort to construct the classical Lie algebras:
\begin{itemize}
  \item the special linear algebra $\mathfrak{sl}(n, R)$,
  \item the (special) orthogonal algebra $\mathfrak{so}(n, R)$,
  \item the symplectic algebra $\mathfrak{sp}(n, R)$,
\end{itemize}
for a finite type $n$ and commutative ring $R$.

These were all constructed as Lie subalgebras of the algebra of square matrices with
entries in $R$. Note that this already includes a design choice: the algebra of $n \times n$
matrices is equivalent, but not equal, to the algebra of endomorphisms of the free module on $n$.
In Mathlib, the free module on $n$ is denoted \texttt{n → R}. We thus also formalised the
equivalence of Lie algebras \extlink{algebra/lie/matrix.lean\#L38}:
\begin{lstlisting}
lie_equiv_matrix' :
  module.End R (n → R) ≃ₗ⁅R⁆ matrix n n R
\end{lstlisting}
so that results could be transported. It is not clear if constructions as matrices or
endomorphisms should be preferred.

\subsection{The special linear algebra}
The Lie subalgebra $\mathfrak{sl}(n, R)$ is the square matrices of trace zero. We thus
added a definition of the trace of a matrix and proved basic properties including:
\begin{align*}
  \tr(AB) = \tr(BA),
\end{align*}
for matrices $A$, $B$ with entries in a commutative semiring. In fact this was deduced
from the following result about transposes which does not assume
commutativity \extlink{linear_algebra/matrix/trace.lean\#L77}:
\begin{lstlisting}
@[simp] lemma trace_transpose_mul
  (A : matrix m n R) (B : matrix n m R) :
  trace n R R (Aᵀ ⬝ Bᵀ) = trace m R R (A ⬝ B) :=
finset.sum_comm
\end{lstlisting}
Note that the proof is a direct invocation of an existing lemma \texttt{finset.sum_comm};
this is unsurprising, Mathlib contains a comprehensive library about finite sums and products.
With the above in hand, it was easy to prove \extlink{algebra/lie/classical.lean\#L76}:
\begin{lstlisting}
@[simp] lemma matrix_trace_commutator_zero
  (A B : matrix n n R) :
  matrix.trace n R R ⁅A, B⁆ = 0 :=
\end{lstlisting}
from which it follows that $\mathfrak{sl}(n, R)$ is indeed a Lie subalgebra.

\subsection{The skew-adjoint algebras}
We already met $\mathfrak{so}(n, R)$ in section \ref{skewSymmSubsect},
it is the subset of matrices $A$ such that:
\begin{align*}
  A^T = -A.
\end{align*}
More generally, by the results of section \ref{skewAdjointSect}, given any square matrix $J$, the
subset of matrices $A$ such that:
\begin{align}\label{AJSkewAdjointEqn}
  A^T J = -J A
\end{align}
form a Lie subalgebra. The remaining classical Lie algebra $\mathfrak{sp}(n, R)$ can be
defined as the subset of $2n \times 2n$ matrices $A$ satisfying
\eqref{AJSkewAdjointEqn} with:
\begin{align*}
  J = \left[\begin{array}{cc}
    0_n & -I_n\\
    I_n & 0_n
  \end{array}\right]
\end{align*}
where $0_n$ is the $n \times n$ zero matrix and $I_n$ is the identity matrix.
Here is the formal definition of the above \extlink{algebra/lie/classical.lean\#L127}:
\begin{lstlisting}
def J : matrix (n ⊕ n) (n ⊕ n) R :=
matrix.from_blocks 0 (-1) 1 0

def sp :
  lie_subalgebra R (matrix (n ⊕ n) (n ⊕ n) R) :=
  skew_adjoint_matrices_lie_subalgebra (J n R)
\end{lstlisting}

However this is not the end of the story: different choices of $J$ yield alternate models of
the classical Lie algebras. These alternate models are different as Lie subalgebras but
equivalent as abstract Lie algebras. For example, the subset of matrices $A$ satisfying
\eqref{AJSkewAdjointEqn} with:
\begin{align}\label{JsoEven}
J = \left[\begin{array}{cc}
  0_n & I_n\\
  I_n & 0_n
\end{array}\right]
\end{align}
is equivalent to $\mathfrak{so}(2n, R)$, and there is a corresponding alternate model for the odd
case $\mathfrak{so}(2n+1, R)$. Furthermore, each model has its advantages\footnote{The main
advantage of the model obtained using \eqref{JsoEven} is that there is a Cartan subalgebra of
diagonal matrices.} so it is important to cater for the various choices of $J$.

We thus also formalized these alternate models, together with relevant proofs of equivalence as
abstract Lie algebras. For example, here is the formal definition of \eqref{JsoEven}
\extlink{algebra/lie/classical.lean\#L215}:
\begin{lstlisting}
def JD : matrix (n ⊕ n) (n ⊕ n) R :=
matrix.from_blocks 0 1 1 0
\end{lstlisting}
and here is the corresponding alternate model \extlink{algebra/lie/classical.lean\#L219}:
\begin{lstlisting}
def type_D :=
skew_adjoint_matrices_lie_subalgebra (JD n R)
\end{lstlisting}
and the statement of its equivalence to a model with a diagonal $J$ matrix
\extlink{algebra/lie/classical.lean\#L262}:
\begin{lstlisting}
noncomputable def type_D_equiv_so'
  [invertible (2 : R)] :
  type_D n R ≃ₗ⁅R⁆ so' n n R :=
\end{lstlisting}

\section{General non-associative algebra and the free Lie algebra}\label{freeAlgebraSect}
We formalised a construction of the free Lie algebra on a type $X$ with
coefficients in a commutative ring $R$. Here is the statement of the universal property
(i.e., left adjointness) as stated in Mathlib with respect to a Lie
algebra $L$ \extlink{algebra/lie/free.lean\#L182}:
\begin{lstlisting}
def lift :
  (X → L) ≃ (free_lie_algebra R X →ₗ⁅R⁆ L) :=
\end{lstlisting}
This definition, and the proof of its universality, was hard-won and is worth comment.

The construction is to take a quotient of the free non-unital, non-associative
algebra\footnote{Strictly speaking the terminology should be {\lq}not-necessarily-unital{\rq} and
{\lq}not-necessarily-associative{\rq} but it is common and easier to say simply
{\lq}non-unital{\rq} and {\lq}non-associative{\rq}.} on $X$ with
coefficients in $R$. We thus needed to define \texttt{free_non_unital_non_assoc_algebra} and
prove its universal property \extlink{algebra/free_non_unital_non_assoc_algebra.lean\#L75}:
\begin{lstlisting}
def lift :
  (X → A) ≃ non_unital_alg_hom R (free_non_unital_non_assoc_algebra R X) A :=
\end{lstlisting}
In the above, $A$ is a general non-unital, non-associative algebra and
\texttt{non_unital_alg_hom} is the type of morphisms of such algebras.

Establishing the above result while adhering to the standards of Mathlib was not straightforward.
The problem was that Mathlib's theories of rings and algebras were entirely specialised to the
unital, associative setting. To handle this without fragmenting the algebraic hierarchy, it was
necessary to insert new classes, notably the typeclass \texttt{non_unital_non_assoc_semiring},
low down in the hierarchy. In a library as large as Mathlib, such changes are significant
undertakings.

Eric Wieser generously took on this
challenge\footnote{This
\href{https://github.com/leanprover-community/mathlib/pull/6786}{pull request}
shows what was required after all other preparatory work had been completed.} and also
showed how to encode a general non-unital,
non-associative algebra:
\begin{lstlisting}
variables {R A : Type*}
  [comm_ring R] [non_unital_non_assoc_semiring A]
  [module R A] [is_scalar_tower R A A]
  [smul_comm_class R A A]
\end{lstlisting}
After Wieser's work (see also \cite{Wieser21}) it was essentially straightforward to
define \texttt{free_non_unital_non_assoc_algebra}
by dropping the assumption of associativity in the
existing monoid algebra construction and proving the corresponding universal property with respect
to a magma $M$ \extlink{algebra/monoid_algebra/basic.lean\#L469}:
\begin{lstlisting}
def lift_magma [has_mul M] :
  mul_hom M A ≃ non_unital_alg_hom R
    (monoid_algebra R M) A :=
\end{lstlisting}
With this in hand, the author was able to make the key definition:
\begin{lstlisting}
def free_non_unital_non_assoc_algebra :=
monoid_algebra R (free_magma X)
\end{lstlisting}
and the corresponding universal property followed trivially.

Finally the free Lie algebra was constructed as a quotient using the following
relation \extlink{algebra/lie/free.lean\#L69}:
\begin{lstlisting}
local notation `lib` :=
free_non_unital_non_assoc_algebra

inductive rel : lib R X → lib R X → Prop
| lie_self (a : lib R X) : rel (a*a) 0
| leibniz_lie (a b c : lib R X) :
    rel (a*(b*c)) (((a*b)*c) + (b*(a*c)))
| smul (t : R) (a b : lib R X) :
    rel a b → rel (t•a) (t•b)
| add_right (a b c : lib R X) :
    rel a b → rel (a+c) (b+c)
| mul_left (a b c : lib R X) :
    rel b c → rel (a*b) (a*c)
| mul_right (a b c : lib R X) :
    rel a b → rel (a*c) (b*c)
\end{lstlisting}
and its universal property followed easily.

It should be noted that the use of \texttt{inductive} above was necessary because Mathlib does not
yet contain a theory of ideals and their quotients for general non-associative algebras. Filling
this gap would improve the construction even further, though the benefit would be slight.

We should say that it would have been easy to establish what we needed without any of the above
work by ignoring most of Mathlib's
algebra library and taking a quotient of an inductively-defined type with a constructor for
every term of $X$ as well as separate constructors corresponding to the scalar action, additive
law, and Lie bracket. We rejected this low-level approach because it would require a significant
quantity of single-use code, because it would be harder to maintain, because the alternative
approach was an opportunity to start developing a general theory of non-associative rings and
algebras, and because this is very unlikely to be the approach that the informal mathematician would take.

We should also say that we rejected an approach that constructs the free Lie algebra as the
smallest Lie subalgebra of the free unital, associative algebra containing the generating type $X$.
This can be expressed in Lean as:
\begin{lstlisting}
lie_subalgebra.lie_span R (free_algebra R X)
  (set.range (free_algebra.ι R))
\end{lstlisting}
This approach is mathematically appealing but the proof that this construction satisfies the
universal property appears to need a powerful version of the Poincar\'e-Birkhoff-Witt theorem
(see \cite{MOFreeLie} as well as \cite{bourbaki1975} I §2.7, §3.1).

\section{The exceptional Lie algebras}\label{exceptionalSect}
We assume for now that the coefficients $R$ form an algebraically closed field of
characteristic zero. The work under discussion does not make this assumption but it will simplify
the discussion here if we do.

There are numerous beautiful ways to construct the five exceptional Lie algebras
$\mathfrak{g}_2, \mathfrak{f}_4, \mathfrak{e}_6, \mathfrak{e}_7, \mathfrak{e}_8$
(e.g., \cite{MR1428422}, \cite{MR0219578}, \cite{MR2140725}, \cite{MR2434742}, \cite{MR3652779}) but the
most useful construction from the point of view of proving the classification theorem
(see section \ref{classificationSect}) is
probably\footnote{The only real competitor being \cite{MR3652779}.}
an approach due to Serre \cite{serre1965}.
This approach takes a square matrix of integers as input and yields a Lie algebra. When the
matrix is the Cartan matrix of a semisimple Lie algebra, we recover the corresponding Lie
algebra, together with a splitting Cartan subalgebra.

If $A$ is an $l \times l$ Cartan matrix, the corresponding Lie algebra is defined to be the
quotient of the free Lie algebra on $3l$ generators:
$H_1, H_2, \ldots H_l$, $E_1, E_2, \ldots, E_l$, $F_1, F_2, \ldots, F_l$
by the following relations:
\allowdisplaybreaks
\begin{align*}
  [H_i, H_j] &= 0\\
  [E_i, F_i] &= H_i\\
  [E_i, F_j] &= 0 \quad\text{if $i \ne j$}\\
  [H_i, E_j] &= A_{ij}E_j\\
  [H_i, F_j] &= -A_{ij}F_j\\
  \ad(E_i)^{1 - A_{ij}}(E_j) &= 0 \quad\text{if $i \ne j$}\\
  \ad(F_i)^{1 - A_{ij}}(F_j) &= 0 \quad\text{if $i \ne j$}\\
\end{align*}
Thanks to the construction of the free Lie algebra described in section \ref{freeAlgebraSect},
it was easy to implement Serre's construction in Mathlib and thus to define the exceptional Lie
algebras. For example, here is Mathlib's definition
of $\mathfrak{f}_4$ \extlink{algebra/lie/cartan_matrix.lean\#L234}:
\begin{lstlisting}
def cartan_matrix.F₄ : matrix (fin 4) (fin 4) ℤ :=
![![ 2, -1,  0,  0],
  ![-1,  2, -2,  0],
  ![ 0, -1,  2, -1],
  ![ 0,  0, -1,  2]]

abbreviation f₄ :=
cartan_matrix.F₄.to_lie_algebra R
\end{lstlisting}

What's more, thanks to Ed Ayers's Lean Widgets \cite{Ayers21}, it was easy to generate the Dynkin
diagram corresponding to a Cartan matrix. For example, here is a screenshot from the author's
proof-of-concept widget, written in Lean, which reads Mathlib's definition of the $E_8$ Cartan
matrix and renders the corresponding Dynkin diagram:

\par\vskip\baselineskip
\begin{center}
  \includegraphics[width=\columnwidth]{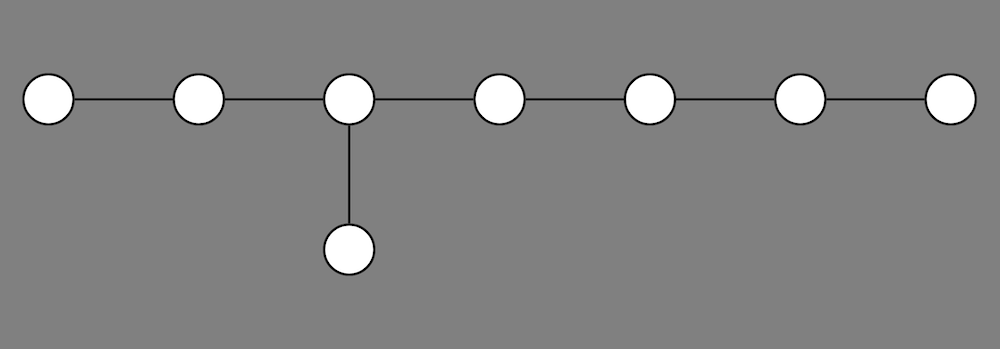}
  \textbf{Mathlib's $E_8$ Dynkin diagram rendered using Lean}
\end{center}
\par\vskip\baselineskip

Finally, we should confess that we have yet to prove almost anything about the exceptional
Lie algebras. The path is clear but much
work remains until we can prove key facts such as their simplicity, or for example:
\begin{lstlisting}
def dimension_g₂ : Prop := finrank (g₂ ℂ) = 14
\end{lstlisting}

\section{Stating the classification}\label{classificationSect}
An important milestone, passed in the course of this work, was teaching Lean the statements
of the classification of semisimple Lie algebras. Using the above work, we defined what
it means for a Lie algebra to be simple \extlink{algebra/lie/semisimple.lean\#L44}:
\begin{lstlisting}
class is_simple extends
  lie_module.is_irreducible R L L : Prop :=
(non_abelian : ¬is_lie_abelian L)
\end{lstlisting}
and likewise what it means to be semisimple \extlink{algebra/lie/semisimple.lean\#L54}:
\begin{lstlisting}
class is_semisimple : Prop :=
(semisimple : radical R L = ⊥)
\end{lstlisting}
Basic related results like the fact that a simple Lie algebra is semisimple
were also proved \extlink{algebra/lie/semisimple.lean\#L76}:
\begin{lstlisting}
instance is_semisimple_of_is_simple
  [h : is_simple R L] : is_semisimple R L :=
\end{lstlisting}

Modulo some boilerplate to establish notation, the classification statements are then:

\begin{lstlisting}
variables (K L : Type*)

/- Let K be an algebraically closed field of
   characteristic zero. -/
variables [field K] [is_alg_closed K]
  [char_zero K]

/- Let L be a finite-dimensional Lie algebra
   over K. -/
variables [lie_ring L] [lie_algebra K L]
  [finite_dimensional K L]

def simple_classification : Prop :=
  is_simple K L ↔
  ((L ≅ₗ⁅K⁆ g₂ K) ∨
   (L ≅ₗ⁅K⁆ f₄ K) ∨
   (L ≅ₗ⁅K⁆ e₆ K) ∨
   (L ≅ₗ⁅K⁆ e₇ K) ∨
   (L ≅ₗ⁅K⁆ e₈ K) ∨
   (∃ l, (L ≅ₗ⁅K⁆ sl n K) ∧ 1 < n) ∨
   (∃ l, (L ≅ₗ⁅K⁆ sp n K) ∧ 2 < n) ∨
   (∃ l, (L ≅ₗ⁅K⁆ so n K) ∧ 4 < n ∧ n ≠ 6))

def semisimple_classification : Prop :=
  is_semisimple K L ↔
  ∃ n (I : fin n → lie_ideal K L),
    (L ≅ₗ⁅K⁆ (⨁ i, I i)) ∧ ∀ i, is_simple K (I i)
\end{lstlisting}

Note that \texttt{simple_classification} contains several of the
{\lq}{\lq}exceptional isomorphisms{\rq}{\rq}. E.g.,
$\mathfrak{so}(6)$ is simple so it must be isomorphic to one of the other algebras on the list. For
dimensional reasons, this has to be $\mathfrak{sl}(4)$. Likewise for the other cases excluded.

Note also that if we pursue a proof of the classification, we will probably restate
\texttt{simple_classification} in terms of algebras
constructed from Cartan matrices of types $A$, $B$, $C$, $D$ rather than the models
$\mathfrak{sl}$, $\mathfrak{sp}$, $\mathfrak{so}$ defined in terms of matrices.

\section{Weight spaces and root spaces}\label{weightRootSect}
Just as a key tool when studying the behaviour of a linear operator is to decompose the space on
which it acts into a sum of (generalised) eigenspaces, a key tool when studying a Lie module $M$
of Lie algebra $L$ is to decompose $M$ into a sum of simultaneous eigenspaces of $x$, as $x$ ranges
over $L$. These simultaneous generalised eigenspaces are known as the weight spaces of $M$.

When $L$ is nilpotent, it follows from the binomial theorem that weight spaces are Lie submodules.
Even when $L$ is not nilpotent, it may be useful to study its Lie modules by restricting them
to a nilpotent subalgebra (e.g., a Cartan subalgebra). In the particular case when we view $L$ as a
module over itself via the adjoint action, the weight spaces of $L$ restricted to a nilpotent
subalgebra are known as root spaces.

We formalised these concepts in Lean as follows \extlink{algebra/lie/weights.lean\#L67}:
\begin{lstlisting}
def pre_weight_space (χ : L → R) :
  submodule R M :=
⨅ (x : L), (to_endomorphism R L M x).
  maximal_generalized_eigenspace (χ x)
\end{lstlisting}
and \extlink{algebra/lie/weights.lean\#L176}:
\begin{lstlisting}
def weight_space [lie_algebra.is_nilpotent R L] (χ : L → R) : lie_submodule R L M :=
{ lie_mem := ...
  .. pre_weight_space M χ }
\end{lstlisting}
and finally \extlink{algebra/lie/weights.lean\#L243}:
\begin{lstlisting}
abbreviation root_space (H : lie_subalgebra R L)
  [lie_algebra.is_nilpotent R H] (χ : H → R) :
  lie_submodule R H L := weight_space L χ
\end{lstlisting}

There is actually quite a lot going on above. For one thing, the definition of \texttt{root_space}
requires Lean to recognise that we can regard $L$ as a Lie module over $H$. This is achieved (in part)
via the following typeclass instance registered far away in the Lie subalgebra
theory \extlink{algebra/lie/subalgebra.lean\#L148}:
\begin{lstlisting}
instance (H : lie_subalgebra R L) :
  lie_module R H M :=
\end{lstlisting}
This is a good example of typeclasses working well: the informal mathematician would not waste
space being explicit about details like this here, and thanks to the typeclass system, the
formal mathematician need not do so either.

More significantly, the proof of \texttt{lie_mem} which we have omitted in the above
definition of \texttt{weight_space} is not quite trivial. The key step is the following lemma
(applied with $\chi_1 = 0$ and $\chi_2 = \chi$) \extlink{algebra/lie/weights.lean\#L161}:
\begin{lstlisting}
lemma
  lie_mem_pre_weight_space_of_mem_pre_weight_space
  {χ₁ χ₂ : L → R} {x : L} {m : M}
  (hx : x ∈ pre_weight_space L χ₁)
  (hm : m ∈ pre_weight_space M χ₂) :
  ⁅x, m⁆ ∈ pre_weight_space M (χ₁ + χ₂) :=
\end{lstlisting}
The proof is similar (though not quite the same) as the proof that if $a, b$ are two commuting
nilpotent elements of a semiring, then their sum $a + b$ is nilpotent. The standard proof of this
is to apply the binomial theorem. In our case, for each element of $L$ we obtain commuting
elements of $\End (L \otimes M)$ and again the proof is to apply the binomial theorem for this
ring. Happily, tensor products, and a general version of the binomial theorem had already been
formalised in Mathlib, so we could just appeal to this theory.

The function $\chi$ appearing in these definitions is the candidate family of eigenvalues,
and is said to be a weight or root when the corresponding weight space or root space is
non-empty \extlink{algebra/lie/weights.lean\#L224}:
\begin{lstlisting}
def is_weight : Prop := weight_space M χ ≠ ⊥
\end{lstlisting}
and \extlink{algebra/lie/weights.lean\#L251}:
\begin{lstlisting}
abbreviation is_root := is_weight H L
\end{lstlisting}

Weights and roots are the start of a sizeable branch of Lie theory. Various foundational
results such as \extlink{algebra/lie/weights.lean\#L294}:
\begin{lstlisting}
def root_space_weight_space_product
  (χ₁ χ₂ χ₃ : H → R) (hχ : χ₁ + χ₂ = χ₃) :
  (root_space H χ₁) ⊗[R]
  (weight_space M χ₂) →ₗ⁅R,H⁆
  weight_space M χ₃ :=
\end{lstlisting}
are in place but much work remains to be done.

\section{Final words}\label{finalWordsSect}
\subsection{Trivial proofs should be trivial}
When building a library the size of Mathlib, one must constantly try to be mindful of how one's
work will scale as more is built upon it. One metric for the health of a particular area of the
library is how much effort one is forced to put into proving trivialities. We share an example
of what this looks like when things go well.

Given a type $X$ and a commutative ring $R$ one can use this data to build the
free unital, associative algebra $A(R, X)$. However, there is another way to build a
unital, associative algebra from this data: one first builds the free Lie algebra $L(R, X)$
and then takes its universal enveloping algebra $U(L(R, X))$. A simple diagram chase reveals
that these are the same, in particular:
\begin{align*}
  U(L(R, X)) \simeq A(R, X).
\end{align*}

Mathlib knows this fact; here is the proof (using some notational shortcuts for readability)
\extlink{algebra/lie/free.lean\#L227}:
\begin{lstlisting}
def universal_enveloping_equiv_free_algebra :
  universal_enveloping_algebra R (free_lie_algebra R X) ≃ₐ[R]
  free_algebra R X :=
alg_equiv.of_alg_hom
  (liftu R $ liftl R $ ιa R)
  (lifta R $ (ιu R) ∘ (ιl R))
  (by { ext, simp, })
  (by { ext, simp, })
\end{lstlisting}
The point of the above is the two lines that read \texttt{(by \string{ ext, simp, \string})}. This is
the Lean code discharging the proof obligations which correspond to the informal
mathematician's diagram chase. It is encouraging that they are trivial applications of
standard tactics.

\subsection{The Lie algebra of a Lie group}
Far away in a different corner of Mathlib, S\'ebastien Gou\"ezel has developed a theory of
differentiable manifolds. Building on top of this, Nicol\`o Cavalleri, under the supervision
of Anthony Bordg, has defined Lie groups and has used it to construct the Lie algebra associated
to a Lie group \cite{BordgCavalleri21}.

\subsection{A specific example}
In section \ref{ubiquitySubsect} we motivated the formalisation of Lie algebras by highlighting
areas of mathematics where they appear. However there also exist specific examples where their
formalisation would help directly. A striking case is the recent paper of Le Floch and Smilga
\cite{MR3851538}. This is a pure mathematics paper in which an interesting abstract problem is
reduced to a finite calculation. The problem was settled by running an algorithm on a
computer, and the authors used SageMath, LiE \cite{lieSoftware2000}, and a custom Coq
program\footnote{The code for this is available at
\href{https://github.com/blefloch/lie-algebra-w0}{https://github.com/blefloch/lie-algebra-w0}.}
written specifically for their purpose.

It is not hard to imagine developing the Lie theory library described here into a platform upon
which calculations such as that of Le Floch and Smilga could be run, and formally verified.

\subsection{Proof of classification}
With the statement of the classification theorem formalised, it is tempting to consider
formalising its proof. Several key concepts such as Cartan subalgebras, weight spaces, and root
spaces have also been formalised. The evidence so far is that formalising a proof of the
classification would be non-trivial but is absolutely within reach.

\section*{Acknowledgements}
It is a pleasure to acknowledge the significant help the author received from numerous members of
the thriving Mathlib community. Almost all of the
\href{https://leanprover-community.github.io/meet.html#maintainers}{maintainers} were of
direct assistance at some point. Special thanks are owed to Johan Commelin and Eric Wieser for
their astonishing appetite to review pull requests (and excellent suggestions) as well as to
Scott Morrison for providing the
motivation to take up this work and for writing the \texttt{noncomm_ring} tactic. I am also
grateful to Kevin Buzzard for frequent advice, guidance, and encouragement.

\bibliographystyle{ACM-Reference-Format}
\bibliography{report}

\end{document}